\let\useblackboard=\iftrue
%
%
\newfam\black
\input harvmac.tex
\noblackbox

\def\Title#1#2{\rightline{#1}
\ifx\answ\bigans\nopagenumbers\pageno0\vskip1in%
\baselineskip 15pt plus 1pt minus 1pt
\else
\def\listrefs{\footatend\vskip 1in\immediate\closeout\rfile\writestoppt
\baselineskip=14pt\centerline{{\bf References}}\bigskip{\frenchspacing%
\parindent=20pt\escapechar=` \input
refs.tmp\vfill\eject}\nonfrenchspacing}
\pageno1\vskip.8in\fi \centerline{\titlefont #2}\vskip .5in}

\ifx\answ\bigans\def\tcbreak#1{}\else\def\tcbreak#1{\cr&{#1}}\fi
\useblackboard
\message{If you do not have msbm (blackboard bold) fonts,}
\message{change the option at the top of the tex file.}
\font\blackboard=msbm10 scaled \magstep1
\font\blackboards=msbm7
\font\blackboardss=msbm5
\textfont\black=\blackboard
\scriptfont\black=\blackboards
\scriptscriptfont\black=\blackboardss

\else

\fi
\def\yboxit#1#2{\vbox{\hrule height #1 \hbox{\vrule width #1
\vbox{#2}\vrule width #1 }\hrule height #1 }}
\def\fillbox#1{\hbox to #1{\vbox to #1{\vfil}\hfil}}
\def\ybox{{\lower 1.3pt \yboxit{0.4pt}{\fillbox{8pt}}\hskip-0.2pt}}

\def\comments#1{}

\def\II{\relax{I\kern-.07em I}}

\def\IZ{\relax\ifmmode\mathchoice
{\hbox{\cmss Z\kern-.4em Z}}{\hbox{\cmss Z\kern-.4em Z}}
{\lower.9pt\hbox{\cmsss Z\kern-.4em Z}}
{\lower1.2pt\hbox{\cmsss Z\kern-.4em Z}}\else{\cmss Z\kern-.4em
Z}\fi}
\def\IB{\relax{\rm I\kern-.18em B}}
\def\IC{{\relax\hbox{$\inbar\kern-.3em{\rm C}$}}}
\def\ID{\relax{\rm I\kern-.18em D}}
\def\IE{\relax{\rm I\kern-.18em E}}
\def\IF{\relax{\rm I\kern-.18em F}}
\def\IG{\relax\hbox{$\inbar\kern-.3em{\rm G}$}}
\def\IGa{\relax\hbox{${\rm I}\kern-.18em\Gamma$}}
\def\IH{\relax{\rm I\kern-.18em H}}
\def\II{\relax{\rm I\kern-.18em I}}
\def\IK{\relax{\rm I\kern-.18em K}}
\def\IP{\relax{\rm I\kern-.18em P}}

\font\cmss=cmss10 \font\cmsss=cmss10 at 7pt
\def\IR{\relax{\rm I\kern-.18em R}}

\def\IZ{\relax\ifmmode\mathchoice
{\hbox{\cmss Z\kern-.4em Z}}{\hbox{\cmss Z\kern-.4em Z}}
{\lower.9pt\hbox{\cmsss Z\kern-.4em Z}}
{\lower1.2pt\hbox{\cmsss Z\kern-.4em Z}}\else{\cmss Z\kern-.4em
Z}\fi}
\def\IB{\relax{\rm I\kern-.18em B}}
\def\IC{{\relax\hbox{$\inbar\kern-.3em{\rm C}$}}}
\def\ID{\relax{\rm I\kern-.18em D}}
\def\IE{\relax{\rm I\kern-.18em E}}
\def\IF{\relax{\rm I\kern-.18em F}}
\def\IG{\relax\hbox{$\inbar\kern-.3em{\rm G}$}}
\def\IGa{\relax\hbox{${\rm I}\kern-.18em\Gamma$}}
\def\IH{\relax{\rm I\kern-.18em H}}
\def\II{\relax{\rm I\kern-.18em I}}
\def\IK{\relax{\rm I\kern-.18em K}}
\def\IP{\relax{\rm I\kern-.18em P}}

\font\cmss=cmss10 \font\cmsss=cmss10 at 7pt
\def\IR{\relax{\rm I\kern-.18em R}}

\def\tilde{\widetilde}
\def\frac#1#2{{{#1} \over {#2}}}

\Title{\vbox{\baselineskip12pt\hbox{hep-th/9712166}
\hbox{SLAC-PUB-7721, SCIPP-97/36}}}
{\vbox{\centerline{New M-theory Backgrounds with Frozen Moduli }
\centerline{}}}
\centerline{Michael Dine}
\centerline{dine@scipp.ucsc.edu}
\centerline{Santa Cruz Institute for Particle Physics}
\centerline{University of California}
\centerline{Santa Cruz, CA   95064}
\centerline{and}
\centerline{Eva Silverstein}
\centerline{evas@slac.stanford.edu}
\centerline{Stanford Linear Accelerator Center}
\centerline{Stanford University}
\centerline{Stanford, CA 94309, USA}
\bigskip
\noindent

We propose examples, which involve orbifolds
by elements of the U-duality group, with M-theory moduli fixed at
the eleven-dimensional Planck scale.
We begin by  reviewing asymmetric
orbifold constructions in perturbative string theory, which
fix radial moduli at the string scale.  Then
we consider non-perturbative aspects of those backgrounds
(brane probes and the orbifold action from the eleven-dimensional
point of view).  This leads us to consider mutually non-perturbative
group actions.  Using a combination of dualities, matrix theory,
and ideas for the
generalization of the perturbative orbifold prescription, we
present evidence
that the examples we construct are consistent M-theory backgrounds.
In particular we argue that there should be consistent non-supersymmetric 
compactifications of M-theory.

\Date{December 1997}

\lref\mukhi{K. Dasgupta and S. Mukhi, ``F Theory at Constant Coupling'',
Phys. Lett. B385: 125-131, 1996 
hep-th/9606044.}
\lref\keith{K. Dienes, Nucl. Phys. B429: 533-588, 1994, hep-th/9402006;
 K. Dienes, M. Moshe, and R. Myers,
Phys. Rev. Lett. 74: 4767-4770, 1995, hep-th/9503055.} 

\lref\otherns{S. Kachru and E. Silverstein,
Nucl. Phys. B463: 369-382, 1996, hep-th/9511228;
O. Bergman and M. R. Gaberdiel, hep-th/9701137;
J. Blum and K. Dienes, hep-th/9707160 and hep-th/9707148.}
\lref\kv{S. Kachru and C. Vafa, ``Exact Results for N=2 Compactifications
of Heterotic Strings'', {\it Nucl. Phys.} {\bf B} 450: 69-89  (1995),
hep-th/9505105.}
\lref\edthree{E. Witten, ``Strong Coupling and the 
Cosmological Constant'',
Mod.Phys.Lett. A10: 2153-2156, 
(1995) hep-th/9506101.}
\lref\moore{A.Losev, G. Moore, and S. Shatashvili,  
``M\& m's'' hep-th/9707250.}
\lref\brs{M. Berkooz, M. Rozali, and N. Seiberg, 
``Matrix Description of M theory on $T^4$ and $T^5$'',
Phys.Lett.B408:105-110,1997, hep-th/9704089.}
\lref\seinew{N. Seiberg, ``New Theories in Six Dimensions and
Matrix Theory on $T^5$ and $T^5/Z_2$'',
Phys.Lett.B408:98-104,1997, hep-th/9705221.}
\lref\vafalevel{C. Vafa, ``Modular Invariance and Discrete Torsion
on Orbifolds'', Nucl.Phys.B273:592, 1986.}
\lref\dsprob{M. Dine and N. Seiberg, ``Couplings and Scales in 
Superstring Models'', Phys. Rev. Lett. 55 (1985) 366.}
\lref\nsv{K. Narain, Sarmadi, and C. Vafa, ``Asymmetric Orbifolds'',
Nucl. Phys. B288 (1987) 551.}
\lref\eddyn{E. Witten, ``String Theory Dynamics in Various 
Dimensions'' Nucl.Phys. B443: 85-126, 1995, hep-th/9503124.}
\lref\sen{A. Sen, ``D0-branes on $T^n$ and Matrix Theory'',
hep-th/9709220.}
\lref\natimm{N. Seiberg, ``Why is the Matrix Model Correct?'',
Phys.Rev.Lett. 79: 3577-3580, 
1997, hep-th/9710009.}
\lref\quasi{J. Harvey, G. Moore
and C. Vafa, ``Quasicrystalline Compactification'',
Nucl.Phys. B304: 269-290, 1988.}
\lref\dlp{J. Dai, R. Leigh, and J. Polchinski, ``New Connections 
between
String Theories'', Mod.Phys.Lett. A4: 2073, 1989.}
\lref\ALE{M. Douglas and G. Moore, ``D-branes, Quivers,
and ALE Instantons'', hep-th/9603167;
M. Douglas, ``Enhanced Gauge Symmetry in M(atrix) Theory'',
hep-th/9612126.}
\lref\orb{L. Dixon, J. Harvey, C. Vafa, and E. Witten,
``Strings on Orbifolds'', Nucl.Phys. B274: 285-314, 1986;
Nucl.Phys. B261: 678-686, 1985.}
\lref\maldstrom{J. Maldacena and A. Strominger, 
``Semiclassical Decay of Near-Extremal Fivebranes'',
hep-th/9710014.}
\lref\chs{C. Callan, J. Harvey, and A. Strominger, 
``Supersymmetric String Solitons'', 
In *Trieste 1991, Proceedings, String theory and 
quantum gravity '91* 208-244 and
Chicago Univ. - EFI 91-066 
(91/11,rec.Feb.92) 42 p., hep-th/9112030 }
\lref\vafaqsymm{C. Vafa, ``Quantum Symmetries of String Vacua'',
Mod.Phys.Lett. A4: 1615,1989.}
\lref\Osixteen{L. Alvarez-Gaume, P. Ginsparg, G. Moore, and C. Vafa,
``An $O(16)\times O(16)$ Heterotic String'', 
Phys. Lett. B171 (1986) 155; L. Dixon and J. Harvey,
``String Theories in Ten Dimensions Without Space-time Supersymmetry'',
Nucl. Phys. B274 (1986) 93.}
\lref\fs{W. Fischler and L. Susskind, ``Dilaton Tadpoles, String Condensates,
and Scale Invariance'', Phys.Lett. B171: 383, 1986; 
Phys. Lett. B173: 262, 1986.}
\lref\uman{A. Kumar and C. Vafa, ``U-manifolds'', 
Phys. Lett. B396: 85-90, 1997,
hep-th/9611007.}
\lref\kutsei{D. Kutasov and N. Seiberg, ``Number of Degrees of
Freedom, Density of States, and Tachyons in String Theory
and CFT'', Nucl. Phys. B358: 600-618, 1991.} 
\lref\tom{T. Banks, ``Matrix Theory'', hep-th/9710231.}
\lref\senorb{A. Sen, ``Duality and Orbifolds'', 
Nucl. Phys. B474: 361-378, 1996, hep-th/9604070.}

\newsec{Introduction}

One of the most interesting issues in M theory is the question of how
the moduli become fixed.  The natural length scale for the
various radii is the eleven-dimensional
Planck scale, $l_P$.  Generic geometrical M-theory backgrounds
preserving supersymmetry have either a moduli space of vacua, or
develop a superpotential which vanishes at a supersymmetric solution
which exists at infinity in some direction in moduli space \dsprob.

In string theory, a set of non-geometrical backgrounds was introduced
in \nsv\ in which many moduli are projected out from the start.
This rather economical method can eliminate radial moduli, though
not the dilaton, in string theory.\foot{One can eliminate the
dilaton in compactifications down to {\it two dimensions} \quasi,
but in four dimensions the dilaton vertex operator remains invariant
under perturbative string orbifolds that preserve
the $4d$ Lorentz group.}  The radii become fixed
at the string scale, $l_S$.   
In this paper we study how these compactifications work
non-perturbatively.  We go on
to argue that it is possible to generalize these
constructions to orbifolds
in M theory which freeze moduli at their natural scale, $l_P$.
In the simplest example of this kind, the orbifold
group breaks all the supersymmetry.  Thus the perturbative
problem that non-supersymmetric models are unstable (i.e. develop a dilaton
tadpole)
may
be overcome.

The basic idea behind the asymmetric orbifold construction of
\nsv\ is as follows.  String backgrounds, such as tori, have
discrete symmetries, such as T-duality.  
At generic points in the moduli space, the
symmetry is broken, but at special points it is restored.  At these
points,
one can
orbifold by this symmetry (perhaps combined with other
symmetries of the system) as long as level-matching constraints
are satisfied.

It is interesting to then consider non-perturbative aspects
of the physics of these backgrounds.
Because we know how T-duality acts on the various branes in the
theory, we can determine how the orbifold group acts on the
non-perturbative spectrum (at least the BPS spectrum).  

The orbifold acts differently on left and right-movers on
the string worldsheet.  This, as well as the fact that the radii
are fixed at the string scale, suggest that these backgrounds are
not geometrical \nsv.  It is interesting to consider then what the
moduli spaces of brane probes look like in these theories.  
We find that the branes do have non-trivial moduli spaces.  

These backgrounds fix radial moduli in string theory.  As for the 
problem of fixing the dilaton,
string-string duality (or U-duality) suggests an answer:  
construct an orbifold by S-duality at the self-dual coupling.
Modding out by U-duality symmetries was discussed in 
\mukhi\uman, where a number of interesting examples
can be found.

Four-dimensional string-string duality maps T-duality
to S-duality \eddyn.  So the existence of a theory obtained by
modding out by T-duality on one side implies that there is a
sensible theory obtained by modding out by S-duality on the
dual side (since there is only one theory involved, which happens to
have two dual descriptions).  Roughly speaking, modding out
by both S and T dualities should fix all the moduli.

Another motivation for studying these somewhat exotic compactifications
is matrix theory.  For ordinary toroidal backgrounds,
the matrix description one derives at
finite discrete light cone momentum $N$ does not decouple from
gravity when the background has $\ge 6$ compact dimensions \sen\natimm. 
It will be interesting to see what the situation is for
non-geometrical backgrounds such
as those discussed here.

This paper is organized as follows.  In \S2\ we review asymmetric orbifolds
and discuss branes and their moduli spaces on these backgrounds.
This leads us to try to develop a
more abstract formulation of orbifold theories than
that which was developed for the perturbative string limits.  
In \S3\ we present an
example in which we orbifold by two mutually non-perturbative symmetries,
fixing the moduli and breaking supersymmetry.  We use the
matrix theory formulation of the orbifolds to argue for consistency.
In \S4\ we give a preliminary discussion of the low-energy physics
of these models, and in \S5\ we conclude by discussing several
interesting open issues.

For other discussions of duality and supersymmetry breaking,
see \otherns.

\newsec{Branes and Asymmetric Orbifolds}

Let us consider the following asymmetric orbifold in the perturbative
type IIA string theory on $T^4$.  Take a square torus with
radii 

\eqn\fixR{R_1=R_2=R_3=R_4=l_S} 
and no $B$ field.  
Then we can mod out by a symmetry generated by

\eqn\gaction{g: (x^1_L,x^1_R;x^2_L,x^2_R; x^3_L,x^3_R;x^4_L,x^4_R)\to
(-x^1_L,x^1_R;-x^2_L,x^2_R;-x^3_L,x^3_R;-x^4_L,x^4_R)} 
acting on the left and right moving bosons on the string world sheet.
The action on the RNS fermions is determined by worldsheet 
supersymmetry.  In this model as it stands,
half of the left-moving supersymmetries in the
untwisted sector are projected out, but the supersymmetry
returns in the twisted sector.
But if we combine this symmetry with an action
$(-1)^{F_R}$,
then one obtains no supersymmetry from the twisted sector.

>From the expressions for the left and right moving momenta
(zero modes of $x^i_L,x^i_R, i=1,\dots,4$)

\eqn\pL{p^i_L={m^i\over R}-n^i{R\over l_S^2}}
\eqn\pR{p^i_R={m^i\over R}+n^i{R\over l_S^2}}
we see that the symmetry \gaction, at the self-dual
radii \fixR, exchanges winding number $n$ and momentum number $m$.
The orbifold is then 
a modding out by T-duality, combined with additional action
on fermionic degrees of freedom.
One can compute the complete perturbative string spectrum following
the methods in \nsv.

Let us consider the spectrum of branes in
this background.  For that we simply need to consider the action
of T-duality on the branes.  For D-branes, T-duality exchanges
Dirichlet with Neumann boundary conditions for the open strings
living on their worldvolumes \dlp.  So for example a D0-brane
turns into a D4-brane wrapped on the $T^4$.
The invariant states will then consist of $k$ D0-branes
and $k$ D4-branes.

How does this all look in eleven dimensions?
The D0-brane is a momentum mode $p_{11}$ in the
eleventh dimension, and the D4-brane is a longitudinal
M-5-brane.  So the orbifold exchanges momentum and
winding in the eleventh dimension as well as in
$x^1,\dots,x^4$!  
The constraint on the moduli, \fixR, is
\eqn\fixRM{R_i={l_P^{3\over 2}\over R_{11}^{1\over 2}}.}
So the M-5-brane wrapped on $x^1,\dots,x^4,x^{11}$ indeed
has the same energy, $1/R_{11}$, as the momentum mode
$p_{11}$.

Let us now consider the moduli spaces of these branes.
First consider the untwisted sector.
The 0-0 strings map to 4-4 strings, while
the 0-4 strings map to 4-0 strings.  The positions of the
D0-branes map to Wilson lines in the D4-brane field theory.  
So the combined D0/D4-brane bound state still has a moduli
space whose Coulomb branch is $k$ copies of the torus.
In addition, there are Higgs branches in which the 0-4 and
4-0 strings get VEVs.

This is rather analogous to what happens for D-brane states
on symmetric orbifolds.  For example, consider D0-branes on
the symmetric orbifold $R^4/Z_2$ \ALE.  There one introduces
``mirror'' D0-branes at the $Z_2$-reflected points on $R^4$.
There is then a branch of the moduli space which is just  
$R^2/Z_2$.  In addition, there is
another branch which emanates from the orbifold fixed point.
When the mirror pair of D0-branes sits there, they can separate
in the transverse directions without spoiling the $Z_2$ symmetry.
This branch is related to the twisted states which
live at the orbifold fixed point.    
The ``twisted states'' of the orbifold correspond here to
bound states of the D0-branes stuck at the orbifold fixed point.   
In our case, the ``mirror'' of the D0-brane is the D4-brane.

In the asymmetric case we are considering, there are also
extra open string sectors, analogous to the Ramond and
Neveu-Schwarz sectors one has in imposing the GSO projection.
These sectors yield new open string moduli replacing those
that were projected out from the untwisted sector.  
(This had to happen in the case of the orbifold by \gaction\ without
the additional $(-1)^{F_R}$ action, since this model is equivalent
to the original unorbifolded theory.  It also happens to be
true for the theory with the extra $(-1)^{F_R}$ action as well.) 

\subsec {$3d \to 4d$?}

One might stop at this point and consider the strong-coupling
limit of the perturbative string asymmetric orbifold as a
way to fix the moduli even in M-theory, by taking $R_{11}$
to be the radius of the $4th$ dimension (i.e. by considering
an asymmetric orbifold of IIA on $T^7$).  This may be
related to the proposal of \edthree\ for supersymmetry breaking.
In particular, by taking appropriate combinations of 
the action \gaction, shifts, and $(-1)^{F_R}$ 
on the $T^7$, one can construct examples with $3d$ $N=1$ supersymmetry,
whose strong coupling (four-dimensional) limit may have no supersymmetry.
One example of such an orbifold group is generated by the
following elements acting on the $T^7$:
\bigskip
\vbox{\settabs 3 \columns
\+$a_1$&$a_2$&$a_3$\cr
\+(-1,1)&shift&(-1,1)\cr
\+(-1,1)&shift&shift\cr
\+(-1,1)&(-1,1)&(-1,1)\cr
\+(-1,1)&(-1,1)&shift\cr
\+shift&(-1,1)&(-1,1)\cr
\+shift&(-1,1)&shift\cr
\+shift&shift&(-1,1)\cr
\+$(-1)^{F_R}$&&\cr}

\noindent The shift here is symmetric between left and
right-movers: it is a shift by half a momentum lattice vector (with
no winding component).  This orbifold level matches in all sectors,
and preserves $3d$ $N=1$ supersymmetry.  There is no twisted supersymmetry
or twisted scalars. 
The physics
is very subtle here, because the orbifold has naively violated
the Lorentz symmetry between the fourth dimension and the other
three, and because the radii of the $T^7$ shrink to zero size 
\fixRM\ in the limit $R_{11}\to\infty$.  There is some 
nonlocality in the physics, since as discussed above momentum
in the eleventh (i.e. fourth) dimension is accompanied by a
wrapped M-5-brane.  We hope to pursue this 
further in future work.

\subsec{Toward a non-perturbative definition of orbifolding}

In ref.
\orb, a precise prescription for constructing
orbifold models 
in perturbative string theory was developed.
Call the orbifold group G.  The rule is
that one keeps all $G$-invariant 
single particle states of the original theory,
and adds in twisted sector states obeying a similar condition.
It is not clear how to formulate
a non-perturbative definition of the orbifold.
In particular, multi-particle states which are invariant under G
are discarded if their single-particle components are not invariant.
Within the framework of string perturbation theory, one can also
give a precise set of rules for interactions.

The fact that it is difficult to give a non-perturbative
definition of the orbifold does not mean that the
orbifold does not make sense at strong
coupling.  In theories with sufficient
supersymmetry, starting from the weak
coupling construction, it follows that a moduli
space exists.  (For non-supersymmetric theories, as always
in string/M theory, the situation is less clear.)
Indeed, certain non-perturbative aspects of 
asymmetric orbifolds are accessible to study,
as we now show.  In the context of string-string dualities,
orbifolding has been studied extensively following the suggestion
in \eddyn; see \senorb\ for a discussion of the rules there.

Let us consider the type IIA string theory orbifolded by G.
In M theory, for each perturbative string state--or more 
accurately, for
each $p_{11}=0$ state--there
must be a set of D0-brane bound states with the same quantum
numbers; i.e. there must exist nonzero-$p_{11}$ modes of each state.
These should arise, as discussed in the previous section, from appropriate
bound states of D0-branes.

So we are led to propose that the spectrum of an M-theory
orbifold consists of all the G-invariant {\it bound states}.
This agrees with the prescription in perturbative string theory
of not keeping all invariant multi-particle states.  It also gives
a prescription for defining the more abstract orbifolds we are
interested in here.  In particular, we can consider the
orbifold \fixR - \fixRM\ at any value of $R_{11}$.

Perturbative string orbifolds must satisfy a set of consistency
conditions imposed by modular invariance \vafalevel.  These
level-matching constraints lead to simple conditions on the
orbifold action.  They are sometimes, but not always, equivalent
to anomaly cancellation, which has so far been the only condition
imposed on M theory orbifolds.  
Level-matching ensures that an orbifold model has a
well-defined perturbative string description.  It is 
possible that in general it is not a constraint, since
one has the possibility of adding space-filling branes
whose moduli can render a model consistent \senorb. (One example is
F-theory in $8d$, where seven branes are included
at points in $T^2/Z_2\sim {\bf P}^1$.)

One way to get a handle on level-matching conditions is to consider
the spectrum of D-branes on an orbifold.  As discussed after
\fixRM, in the case of the action \gaction, 
untwisted states seem to correspond to the D0-brane/D4-brane 
wavefunction supported on the interior of their Coulomb
branch, while we expect the twisted states to correspond to
states localized at the origin of the Higgs and Coulomb
branches, in analogy to the twisted states of symmetric
orbifolds \ALE.  

If we tried to act with -1's on six left-movers, for example, 
instead of four, 
we would find that the perturbative string theory does not level-match.
Correspondingly, the D0-brane/D6-brane system breaks supersymmetry,
the branes repel, and there are no analogues of the ``twisted states''.
So a natural guess is that requiring that each element of the orbifold
group maps branes to other branes which preserve supersymmetry
should ensure that the model is consistent.

In six dimensions, i.e. M theory on $T^5$,
one can use matrix theory to define the
orbifold models.  There the consistency conditions are somewhat cleaner,
since in that case U-duality becomes T-duality of the defining
matrix theory (as we will review below) \brs.  This will give us
one set of models.  We will then move on to consider four-dimensional
examples where matrix theory is no longer helpful but we
can at least ensure that the group elements map branes to other mutually
supersymmetric branes.

\newsec{Fixing the Dilaton:  Mutually Non-perturbative Orbifold Groups}

A perturbative string orbifold does not fix all the M-theory moduli
(at best it relates them to $R_{11}$ as in \fixRM ).  
In this section we will generalize the asymmetric orbifold construction
to construct orbifolds of M theory which fix radii at $l_P$.  
We will first study a six-dimensional example because there we
can use the matrix formulation of \brs\seinew\ to get a handle
on the consistency conditions.  Then we will generalize the construction
to four dimensions.  In \S4 we will discuss aspects of the low energy
physics of the examples, including subtleties pertaining to the question
of stability.

\subsec{$6d$ Example}

Let us begin as in \S2\ with M theory on $T^5$.  In the matrix theory
this is given by the (2,0) supersymmetric $6d$ string theory
of \chs\seinew\ compactified on another five-torus $\tilde T^5$.
This torus has radii $\Sigma_1,\dots,\Sigma_5$, which are related
to the radii $L_1,\dots,L_5$ of the spacetime $T^5$ as
\eqn\matmapI{\Sigma_i={l_P^3\over{RL_i}},~~ i=1,\dots,4}
\eqn\matmapII{\Sigma_5={l_P^6\over{RL_1L_2L_3L_4}}}
\eqn\matmapIII{\tilde M_S^2={{R^2L_1L_2L_3L_4L_5}\over{l_P^9}}}
where $R$ is the longitudinal radius and $\tilde M_S$ is the
string scale of the (2,0) string theory.
This theory was obtained \seinew\ by considering the limit of vanishing
string coupling in the background of a symmetric fivebrane \chs.
There is evidence \maldstrom\ 
that although the theory decouples from gravity,
it includes the full conformal field theory describing
strings propagating on the throat of the solution \chs\ as well as
along the five Poincare-invariant dimensions.

What we will need of this background is the fact that it
has a T-duality symmetry $SO(5,5,Z)$ acting on the moduli
of the $\tilde T^5$.  In general, because this string
theory is strongly coupled, we cannot quantize the strings
in the usual manner of perturbative string theory.  However,
in \maldstrom\ it was observed that there is a regime in which
this theory has weakly coupled strings.  Here we will first
discuss the orbifold action in this regime, where the strings
are weakly coupled, and ensure there that the orbifold satisfies
the level-matching conditions.
In particular, let us consider the following
orbifold group $G$, generated as follows by elements $f$ and $g$, in the 
(2,0) string theory:

\bigskip
\vbox{\settabs 3 \columns
\+f&g&fg\cr
\+(-1,1)&(1,1)&(-1,1)\cr
\+(-1,1)&(1,-1)&(-1,-1)\cr
\+(-1,1)&(1,-1)&(-1,-1)\cr
\+(-1,1)&(1,-1)&(-1,-1)\cr
\+(1,1)&(1,-1)&(1,-1)\cr
\+$(-1)^{F_R}$&$(-1)^{F_L}$&$(-1)^{F_L+F_R}$\cr}

In the string theory, all group elements level-match.
Because there is a weakly coupled regime, this is necessary
for consistency of the model.  
Though this is of course not a proof--we do not
know whether this is sufficient for consistency--we take
it as strong evidence that the model is consistent.

The elements $f$ and $g$ together fix
\eqn\fixsig{\Sigma_1=\Sigma_2=\dots =\Sigma_5={1\over{\tilde M_S}}}
This translates in the spacetime theory into the condition 
\eqn\fixL{L_1=L_2=\dots L_5=l_P.}

In spacetime the orbifold group $G$ acts as follows.
The element $f$ acts as T-duality on $L_1,L_2,L_3,L_4$,
along with $(-1)^F$, 
in the IIA theory with respect to which 
$L_5$ corresponds to the ``eleventh'' dimension.
Similarly the element $g$ acts as T-duality
on $L_2,L_3,L_4,L_5$, along with $(-1)^F$, in the IIA theory with respect
to which $L_1$ corresponds to the eleventh dimension.

This orbifold kills all the supersymmetries.  
We start with a 32-component supercharge $\epsilon$ 
in eleven dimensions.  The element $f$ leaves invariant
half of the spinors satisfying $\epsilon=\Gamma_5\epsilon$
(i.e. left-handed supersymmetries in the IIA theory with
respect to which $L_5$ corresponds to the eleventh dimension).
The element $g$ leaves invariant half of the spinors satisfying
$\epsilon=\Gamma_1\epsilon$.  From the point of view of
the original IIA theory, $\Gamma_1$ changes the chirality
of the spinor, so this condition is incompatible with the supersymmetries
left invariant by $g$.  
We could preserve some supersymmetries, at the cost of
introducing scalars with flat directions in their potential.

Without supersymmetry, there is an issue of whether the
matrix theory has flat directions, at least at distances
greater than $l_P$,
which is required for spacetime to emerge.
This is not yet clear to us, but the following points are
relevant.  
As discussed above, the matrix theory for M theory on $T^5$ has
as an analogue model the theory of $N$ NS fivebranes
on $\tilde T^5$, in the limit $g_S\to 0$ (where
$g_S$ is the string coupling).  The
analogue model for our case is the theory of $N$ NS fivebranes
on $\tilde T^5$ in the IIA theory
modded out by the asymmetric orbifold given above,
in the limit $g_S\to 0$.  That theory has fivebranes at separate
points (as long as the separation is greater than 
$1/\tilde M_S = l_P^2/R$) with no force between them.  This
is because the way the force cancels in the supersymmetric
theory is by cancellations between dilaton, graviton,
and antisymmetric tensor exchange.  All these fields are
projected in by the orbifold, so the asymptotic flat directions
remain.  However, 
this is not sufficient to ensure that we have ordinary gravity.

Another feature of our model is the absence of tachyons, 
and the resulting improved supersymmetry properties
at high mass levels \kutsei\keith.
This may be enough to produce asymptotic flat directions at
the right scale in the
potential \tom, though we need better control over the
fivebrane theory in order to analyze this.

\subsec{$4d$ Example}

We will now consider a $4d$ M-theory background obtained by
orbifolding M-theory on $T^7$, with coordinates
$(x_1,x_2,x_3,x_4,x_{5=11},x_6,x_{7=\tilde{11}})$.  
The orbifold group $H$ is generated by two elements.  The
first, $h_1$, can be most easily described by considering
M-theory on this $T^7$ as a IIA string theory with
respect to which $x_{5=11}$ is the eleventh dimension.
Then $h_1$ acts as T-duality on $x_1,x_2,x_3,x_4$,
combined with $(-1)^{F}$.
Similarly, we take $h_2$ to be T-duality on 
$x_3,x_4,x_{5=11},x_6$ combined with
$(-1)^{F_{}}$ in a $\tilde {IIA}$ theory
in which $x_{7=\tilde {11}}$ is the eleventh dimension.
This action fixes all the radii of the $T^7$ to
be $l_P$.    

In this case we do not have a matrix realization to
work with.  We expect, however, that imposing the condition that
each orbifold group element maps branes to mutually 
supersymmetric branes is likely to lead to consistent
models.  We checked this for the $4d$ model just proposed. 
This is automatic for the elements $h_1$ and $h_2$, so
one just needs to check the products.  
For example, the M-5-brane wrapped on $x_1,x_2,x_3,x_4,x_{5=11}$
maps under $h_2h_1$ to an M-2-brane wrapped on 
$x_{5=11},x_{7=\tilde{11}}$.  These two objects preserve
supersymmetry.

\newsec{Low-energy physics of the models}

What can we say about the spectrum of this theory?
The first important question is whether there are scalars in
the low-energy spectrum.  We have certainly
projected out the untwisted radii, since U-duality is only
a symmetry at the self-dual radii.  There can however in principle
be scalars in the ``twisted sectors'' of our orbifolds.  Note
that while the phrase ``twisted sector" refers to the perturbative
construction, in fact these sectors are distinguished by
discrete quantum numbers.  Such quantum symmetries\vafaqsymm\ are
exact in perturbative orbifolds, and thus
they might be expected to exist in this theory as well. 
If we preserve enough supersymmetry (e.g. by not including
the $(-1)^{F_R}$ actions in our orbifolds), this happens because
there are scalars in the supermultiplets.
With enough supersymmetry, these scalars will have flat directions
in their potential, and may be in general connected to geometrical
models by going out along them, as in some of the examples in \kv.

Without supersymmetry, as in the above examples including the
$(-1)^{F_R}$ actions, we have less control over the orbifold.
In order to determine whether there are scalars in the twisted
sectors, we would need to know the quantum numbers of the bound
states of the orbifold theory's Hamiltonian.
We can choose the orbifold action so as to ensure that the 
fermionic zero modes which are free and decoupled generate
non-trivial representations of the Lorentz group and no scalars.
But in principle the rest of the degrees of freedom could interact in such
a way as to cause the ground states to have different
quantum numbers. 

However, even if there are ``twisted'' scalars, it is
likely that a potential is generated which lifts any flat directions
they would otherwise have.  In particular, the twisted scalars
will be charged under the quantum symmetry, so the orbifold point
will be
an extremum of the potential.  This is to be contrasted with
earlier constructions of non-supersymmetric backgrounds in
weakly-coupled string theory \Osixteen, which, though interesting,
are generically unstable to running off to weak coupling.           

Note also that we did not project out the graviton state.  It
corresponds to a diffeomorphism symmetry in the noncompact dimensions
which remains unbroken by our orbifold action.

We should stress a subtlety here.  
As discussed in \S2.2, modulo the matrix construction
we do not have a complete, non-perturbative
description of the orbifolding procedure.  In 
perturbation theory, the orbifold procedure
is not guaranteed to construct a stationary
solution of string theory.  In particular, there are examples where
twisted moduli
are at a maximum, instead of a minimum, of the potential, and
untwisted massive fields have tadpoles.
A well-known example
of this phenomenon is provided by the compactification of
the $O(32)$ heterotic string on a symmetric orbifold.
In this construction, at order $g^2$, there is
a Fayet-Iliopoulos $D$-term, and at order
$g^4$ there is a dilaton tadpole.  At still higher
orders, one expects to generate tadpoles and curvature
for all untwisted moduli and charged fields (e.g. masses
can be generated at fourth order in the
Fayet-Iliopoulos parameter).  In all known cases, it is
possible to shift some charged field so as to cancel
the $D$-term and restore supersymmetry.  But at the level
of the orbifold procedure, it is not clear why this is true.
In the models we have described, supersymmetry
is completely broken, and there are no small dimensionless
parameters.  A priori, then, we might expect that,
while there are no massless states, there
might be tadpoles for massive fields and that the
true ground state,\foot{It is perhaps worth
recalling that in ordinary weak coupling string theory,
tadpoles for massive  particles are not important
since they are cancelled by small shifts, and
are automatically taken care of by properly
``integrating out;" at strong coupling,
the situation is inevitably more complicated.}
if any, might lie far away and have quite
different properties than those suggested by the orbifold
construction.  

However, it will always be the case that the orbifold point
will be an extremum of the potential.  We find it likely
that for some examples this extremum will be a minimum after
taking into account any tadpoles of massive fields.
In particular, we have presented a matrix formulation in 
\S3.1\ in terms of fivebranes
in a non-supersymmetric string theory with $g\to 0$ and
asymptotic supersymmetry.  
We find these features promising, but unfortunately the strong coupling
at the core of the fivebrane precludes a more detailed analysis
at present.

It is important to note that although we have fixed the moduli at
$l_P$, this does not necessarily imply
that the low-energy effective couplings
are strong.  In a theory with exact electric-magnetic
duality, the gauge couplings are necessarily
large at the self
dual point.  
At such a point, one might
worry that the self-dual value of the bare coupling is preserved
in the effective theory due to degeneracy of electric and
magnetic states.  When one orbifolds by S-duality, as we essentially
do here, this objection is evaded, since the orbifold does not
leave all the independent electric and magnetic states.

Finally, and perhaps most crucially, there is still the question
of whether there is a cosmological constant.  In all known examples of
perturbative string theories without supersymmetry,
there is a non-zero cosmological
constant at one loop.  This might suggest
that in theories without moduli, one should expect
a cosmological constant scaled by $M_p$.  However,
there is an important difference between these cases:
in weakly coupled theories, the 1-loop cosmological constant
is proportional to a 1-loop dilaton tadpole \fs.  The evolution
of the system then tends to drive the cosmological
constant to zero.  In the present case, there is no
such tadpole.
There are
very speculative arguments \ref\bankscosmo{For a discussion,
see T.
Banks, ``SUSY Breaking, Cosmology, Vacuum Selection
and the Cosmological Constant in String Theory,''
hep-th/9601151.}, based
principally on the holographic principle \ref\hologram{L.
Susskind, J.Math.Phys. {\bf 36} {1995} 6377,hep-th/9409089.},
that such a cosmological term would not make sense.
Models of the type we have discussed
here should be a testing ground for these ideas.

If there is a non-zero cosmological
constant, then there may well be
a solution of M theory
of this kind, but there is probably no sense in which one can
speak of a ``low energy theory'' at all.  For example, 
terms in the gravitational action involving high powers of the
curvature, ${\cal R}$, will not be suppressed.
%

\newsec{Conclusions and Open Questions}

We have provided evidence that M theory has a set of backgrounds,
essentially non-perturbative generalizations of asymmetric orbifolds,
in which
the moduli are projected out (fixed at $l_P$).  We are limited
computationally by strong couplings in the construction. 
However, there are many interesting
questions this set of models raises.  We would like a more detailed,
direct understanding of what it means to mod out by S-duality, and
how it works just within quantum field theory.  Similarly it would
be very nice to derive simple and general
consistency conditions analogous
to level-matching constraints in perturbative orbifolds. 
We would also like to pursue the 
low-energy physics of these models, in particular the $3d\to 4d$ model
in \S2.1.

\medskip

\centerline{\bf Acknowledgements}

We would like to thank O. Aharony, T. Banks,
L. Dixon, W. Fischler, S. Kachru, A. Lawrence, 
J. Maldacena, A. Rajaraman, N. Seiberg, S. Shenker, and C. Vafa
for helpful discussions.  The work of 
E.S. is supported by the DOE under contract
DE-AC03-76SF00515; that of M.D. is partially supported
by the Department of Energy.

\listrefs

\end